\documentclass{ifacconf}

\usepackage{graphicx}      % include this line if your document contains figures
\usepackage{natbib}        % required for bibliography
\usepackage{amsmath}
\usepackage{amsfonts}
\usepackage{subfigure}
\usepackage{amsmath,amsfonts}
\usepackage{url}

\begin{document}
\begin{frontmatter}

\title{TS-MPC for Autonomous Vehicle using a Learning Approach \thanksref{footnoteinfo}} 

\thanks[footnoteinfo]{The authors wish to thank the support received by the Spanish national project DEOCS (DPI2016-76493-C3-3-R).}

\author[First,Second]{Eugenio Alcala} 
\author[Third]{Olivier Sename} 
\author[First,Second]{Vicenç Puig}
\author[Second]{Joseba Quevedo}

\address[First]{Institut de Rob\`{o}tica i Inform\`{a}tica Industrial (CSIC-UPC). Carrer Llorens Artigas, 4-6, 08028 Barcelona (email: eugenio.alcala@upc.edu).}
\address[Second]{Supervision, Safety and Automatic Control Research Center (CS2AC) of the Universitat Polit\'{e}cnica de Catalunya, Campus de Terrassa, Gaia Building, Rambla Sant Nebridi, 22, 08222 Terrassa, Barcelona.}
\address[Third]{Univ. Grenoble Alpes, CNRS, Grenoble INP, GIPSA-lab, 38000
Grenoble, France.}

\begin{abstract}                
In this paper, the Model Predictive Control (MPC) and Moving Horizon Estimator (MHE) strategies using a data-driven approach to learn a Takagi-Sugeno (TS) representation of the vehicle dynamics are proposed to solve autonomous driving control problems in real-time.
To address the TS modeling, we use the Adaptive Neuro-Fuzzy Inference System (ANFIS) approach to obtain a set of polytopic-based linear representations as well as a set of membership functions relating in a non-linear way the different linear subsystems.
The proposed control approach is provided by racing-based references of an external planner and estimations from the MHE offering a high driving performance in racing mode.
The control-estimation scheme is tested in a simulated racing environment to show the potential of the presented approaches.
\end{abstract}


\end{frontmatter}
%===============================================================================

\section{Introduction}

%%--> 1º
In recent years, the amount of learning-based applications has increased immensely.
Particularly, in the autonomous driving field, we have witnessed new approaches as the end-end driving where the goal consists on guiding the vehicle by means of using learning algorithms and input sensors data.
In \cite{sallab2017deep}, a deep reinforcement learning framework is proposed that takes raw sensor measurements as inputs and outputs driving actions. In \cite{bojarski2016end}, authors use a Convolutional Neural Network (CNN) to obtain the appropriate steering signal from the images of a single front camera.

Nowadays, from a control point of view, several strategies are starting to use learning tools to improve their capabilities while guaranteeing overall system stability.
We have witnessed an advance in this field reaching learning techniques to adjust controllers, identify some parameters inside models or even control non-linear systems.
In \citet{lefevre2015learning, lefevre2015autonomous}, some solutions for controlling the longitudinal velocity of a car based on learning human behaviour are presented.
Also, a Model Predictive Control (MPC) technique using deep CNN to predict cost functions from the camera input images is developed in \cite{drews2017aggressive}.
In \cite{rosolia2017learning, rosolia2017autonomous, rosolia2019learning}, the authors propose a reference-free iterative MPC strategy able to learn from previous laps information.

%%--> 2º
Most of the last approaches were based on learning some policies to drive the vehicle independently of a physical model. 
In this work, we are interested on learning a realistic and accurate representation of the system dynamics to improve the control performance.
In \cite{kabzan2019learning}, authors use a simple starting vehicle model which is enhanced on-line by learning the model error using Gaussian process regression and measured vehicle data.

%%--> 3º
In paper, we propose the use of ANFIS approach, that is an adaptive neuro-fuzzy inference system, to learn the vehicle model. In the same manner as artificial neural networks, it works as a universal approximator \citep{jang1993anfis}. The main purpose of using ANFIS is to learn an input-output mapping based on input data. This tool has been widely used in other engineering fields \citep{ndiaye2018adaptive, jaleel2019identification}.

%% Contribution:
The main contribution of this work is to model accurately a non-linear system as a structured Takagi-Sugeno (TS) representation of the vehicle by means of using machine learning tools and input data. 
In particular, this paper takes advantage of the properties of ANFIS tool to learn a data-driven TS system which will be later used by a predictive optimal control to solve the driving problem.

%% Structure:
The paper is structured as follows. Section \ref{section:2} presents the testing vehicle used in simulation. 
Section \ref{section:3} details the proposed learning-based method and its main components. 
Section \ref{section:4} formulates the control and estimation problems.
Section \ref{section:5} introduces its application to a case study to assess the methodology, as well as various performance results. Finally, Section \ref{section:6} presents several conclusions about the method suitability.

\section{Testing vehicle} \label{section:2}
\begin{figure}[t]
    	\centering
      	\includegraphics[width=75mm]{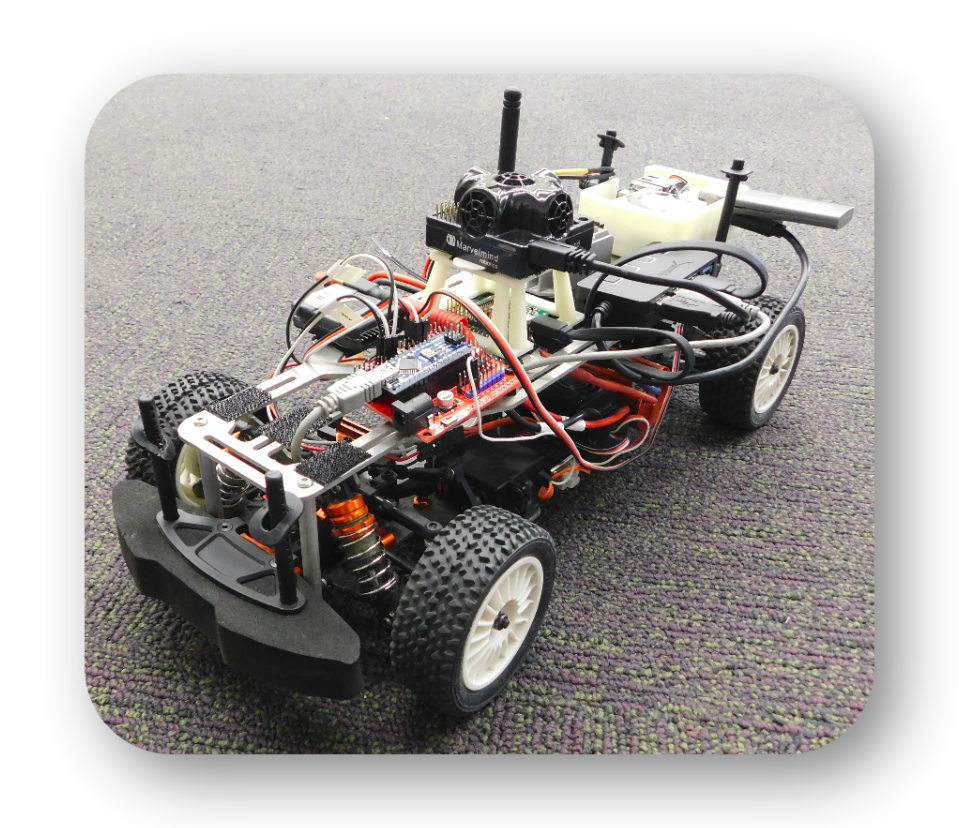}
      	\caption{ Real picture of the vehicle used for simulation}
      	\label{fig:BARC_vehicle}
	\end{figure}

The Berkeley Autonomous Race Car \citep{gonzales2016autonomous} (BARC\footnote{http://www.barc-project.com/}) is a development platform for autonomous driving to achieve complex maneuvers.
This is a 1/10 scale RWD electric remote control (RC) vehicle (see Figure \ref{fig:BARC_vehicle}) that has been modified to operate autonomously.
Mechanically speaking, this has been modified with some decks to protect the on-board electronics and sensors.
%This vehicle includes a basic net of sensors for performing localization. A fusion of IMU, encoders and indoor GPS data is made using an LPV estimator \citep{alcala2018gain} to accurately achieve a good localization while testing.

The non-linear model used in this chapter for simulating the BARC vehicle is presented as
\begin{equation}
	\label{eq:NL_model_for_simulation}
    \begin{aligned}
       % & \dot x = \cos{\theta} v_x - \sin{\theta} v_y \\
       % & \dot y = \sin{\theta} v_x + \cos{\theta} v_y \\
       % & \dot \theta = \omega \\	    
       	& \dot v_x = a_r + \frac{- F_{yf} \sin{\delta} - \mu g}{m} + \omega v_y   \\
        & \dot v_y =  \frac{ F_{yf} \cos{\delta} + F_{yr}}{m} - \omega v_x \\
		& \dot \omega = \frac{ F_{yf} l_f \cos{\delta} - F_{yr} l_r }{I} \\
    	& \alpha_f = \delta - \tan^{-1} \left(  \frac{v_y}{v_x} - \frac{l_f \omega}{v_x} \right) \\
    	& \alpha_r = - \tan^{-1}  \left(  \frac{v_y}{v_x} + \frac{l_r \omega}{v_x} \right) \\    		      
        & F_{yf} = d \sin{ (c \tan^{-1} (b \alpha_f)) } \\
		& F_{yr} = d \sin{ (c \tan^{-1} (b \alpha_r)) } \\        
        %%& F_{df} = \mu m g %+ \frac{1}{2} \rho C_{dAf} v_x^2  
    \end{aligned} \ ,
\end{equation}
\noindent where the dynamic vehicle variables $v_x$, $v_y$ and $\omega$ represent the body frame velocities, i.e. linear in $x$, linear in $y$ and angular velocities, respectively. 
The control variables $\delta$ and $a$ are the steering angle at the front wheels and the longitudinal acceleration vector on the rear wheels, respectively.
$F_{yf}$ and $F_{yr}$ are the lateral forces produced in front and rear tires, respectively. 
This considers the simplified "Magic Formula" model for simulating lateral tire forces where the parameters $b$, $c$ and $d$ define the shape of the curve. 
%These last have been obtained through an identification procedure. 
Front and rear slip angles are represented as $\alpha_f$ and $\alpha_r$, respectively.
$m$ and $I$ represent the vehicle mass and inertia and $l_f$ and $l_r$ are the distances from the vehicle center of mass to the front and rear wheel axes, respectively. $\mu$ and $g$ are the static friction coefficient and the gravity constant, respectively.
All the dynamic vehicle parameters are properly defined in Table \ref{table:vehicle_parameters}.    
\begin{table}[htbp]
    \caption{Dynamic model parameters}
    \label{table:vehicle_parameters}
    \centering
    \begin{tabular}{ l|l||l|l }
    \hline
    Parameter & Value & Parameter & Value \\
    \hline
    \hline
    $l_f$   & 0.125  $m$   &  $l_r$   & 0.125  $m$  \\
    $m$     & 1.98    $kg$ &  $I$     & 0.03 $kg$ $m^2$  \\
    $d$     & 7.76      & $c$   & 1.6 \\
    $b$     & 6.0     & $\mu$ & 0.1  \\ 
    \hline
    \end{tabular}
\end{table}

In this work, with the aim of improving the simulation, Gaussian noise has been introduced in the measured variables as
\begin{equation}
	n_{(\cdot)} \sim N(0, Co_{(\cdot)})
\end{equation}
where $Co_{(\cdot)}$ is the signal covariance. %(see $n_{(\cdot)}$ vector in Figure \ref{fig:schematical_view}).

\section{Learning the TS model} \label{section:3}
In this section, we present the modeling methodology used for obtaining the TS representation of the vehicle dynamic model.
The tool ANFIS \citep{jang1993anfis}, is an adaptive neuro-fuzzy inference machine that is used for learning a particular structure from input-output data.
More in detail, this modeling tool configures a neural network that learns from IO data the dynamic behaviour of the vehicle using back propagation technique and also employing the Recursive Least Squares (RLS) method for adjusting additional parameters.

The methodology consists on providing a dataset to the modeling algorithm (ANFIS). This is composed by vehicle states and inputs that represents a set of particular driving maneuvers guaranteeing rich enough data.
Then, after a learning-based procedure, this provides a set of linear parameters, also known as consequent parameters, and a set of premise parameters or non-linear parameters that define the set of membership functions (MF) that provide the non-linear relationships between the different linear polynomials. One typical membership function is the generalized Gaussian function.

\begin{figure}
	\centering
    \includegraphics[width=85mm]{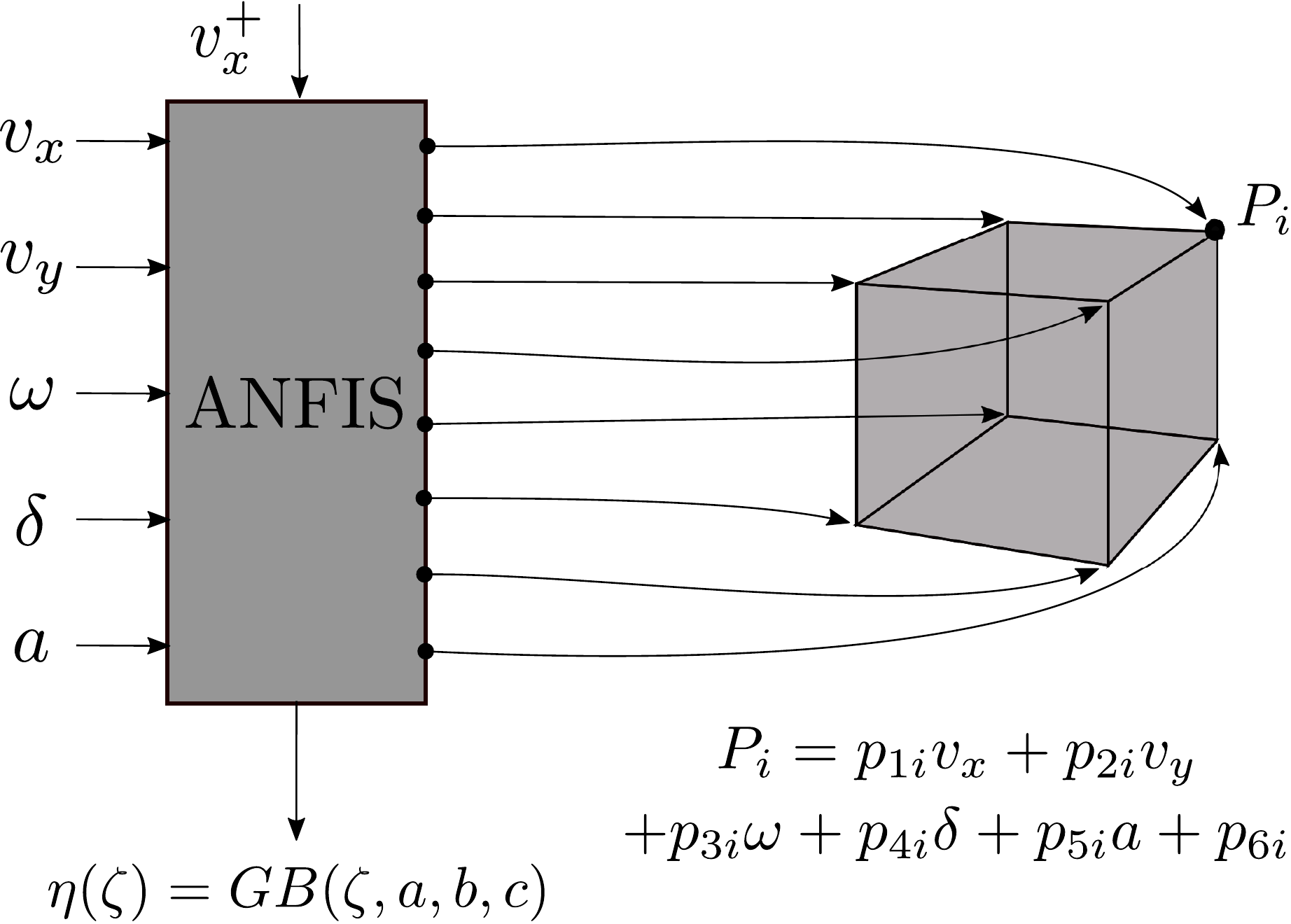}
    \caption{Polytopic TS learning scheme for the $v_x$ sub-system case}
    \label{fig:polytopic_learning_approach}
\end{figure}

However, obtaining the TS representation of a system by means of using this resulting parameters is not trivial. The procedure is based on performing some inverse steps that ANFIS internally does. 
To address this problem we have to follow a set of reformulating steps.
First, due to ANFIS algorithm can be only used for Multi Input Single Output (MISO) systems where just an output variable can be considered. 
Then, we split the system in MISO sub-systems obtaining as many sub-systems as state variables have the system. Our dynamic vehicle model is a third order system so three sub-systems will be obtained and three learning procedures will be carried out.
Once the algorithm has computed conveniently the consequent and premise parameters for each one of the MISO sub-systems, we build the polytopic TS state-space representation for each one of these sub-systems. 
To do this, first, the polynomial representation of each sub-system is formulated as
\begin{equation}
	\label{eq:consequent_parameters}
	\begin{aligned}	
	& P_i = p_{1i}v_x + p_{2i}v_y + p_{3i}\omega + p_{4i}\delta + p_{5i}a + p_{6i} \\
	& \forall i = 1, ..., N_v ,
	\end{aligned}	
\end{equation}
where $P_i$ stands for a linear polynomial representation of the dynamics of a sub-system at a particular states configuration, $p_{ji} , \ \forall j = 1, ..., N_{\zeta}$, are the consequent parameters obtained from ANFIS where $N_{\zeta}$ stands for the number of scheduling variables (See Figure \ref{fig:polytopic_learning_approach}) and $N_v$ represents the number of polytopic vertexes.

\noindent Then, simply by reorganising the terms in \eqref{eq:consequent_parameters} as
\begin{equation}
	P_i = \left[\begin{array}{ccc} p_{1i} & p_{2i} & p_{3i} \end{array}\right]   x + \left[\begin{array}{cc} p_{4i} & p_{5i}\end{array}\right] u + \left[\begin{array}{c} p_{6i} \end{array}\right]
\end{equation}
the polynomial structure is transformed into the discrete-time state-space representation given by
\begin{equation}
	x_i^+ = A_{i} x + B_{i} u + C_{i} \ , \forall i = 1, ..., N_v ,
\end{equation}
where, in order to easy the comprehension from a control point of view, $P_i$ is represented as the sub-system $i$ variable at the next discrete step ($x_i^+$) with symbol $^+$ representing the $k+1$ discrete step.
$A_{i}$, $B_{i}$ and $C_{i}$ define the so-called \emph{vertex systems}, $x = \left[\begin{array}{ccc} v_x & v_y & \omega \end{array}\right]^T$ and $u = \left[\begin{array}{cc} \delta & a \end{array}\right]^T$.

At this point, we use the obtained premise parameters to formulate the membership function. One of the most used is the generalized Gaussian Bell function ($GB$). This is defined by three parameters ($a, b$ and $c$) as follows 
\begin{equation}
	\label{eq:polytopic_membership_function} 
	\begin{aligned}
	& \eta_{m}(\zeta_{o}) = \frac{1}{1 + \frac{\zeta_{o} - c_{mo}}{a_{mo}}^{2 b_{mo}}} , \\
	& \forall m = 1, ..., N_{MF} , \forall o = 1, ..., N_{\zeta} , 
	\end{aligned}
\end{equation}
where $\zeta$ represents the ANFIS input vector of variables (from now on we will refer to them as scheduling variables) and $N_{MF}$ and $N_{\zeta}$ represent the number of MF per scheduling variable and the number of scheduling variables, respectively.
For a common case where $N_{MF}$ is two, then, the normalized weights ($\mu_{N_i}$) are computed following
\begin{equation}
    \mu_i (\zeta) = \prod_{j=1}^{N_{\zeta}}  \xi_{ij}(\eta_{0}, \eta_{1}) , \forall i= 1,...,N_v , \
\end{equation}
where $\xi_{ij}(\cdot)$ corresponds to any of the weighting functions that depend on each rule $i$. Then, using 
\begin{equation}
    \mu_{N_i} (\zeta) = \frac{ \mu_i (\zeta) }{\sum \limits_{j = 1}^{N_v} \mu_j (\zeta)} , \forall i= 1,...,N_v , \
\end{equation}
the normalized weights are obtained.
Note that, each scheduling variable $\zeta_{o}$ is known and varies in a defined interval $\zeta_o \in \left[ \underline{\zeta_o}, \overline{\zeta_o} \right] \in \mathbb{R}$.
Finally, the polytopic TS model for each sub-system is represented as
\begin{equation}
	\label{eq:TS_polytopic_SUBsystem}
	x_j^+ = \sum \limits_{i = 1}^{N_v} \mu_{N_{ji}}(\zeta) (A_{ji} x + B_{ji} u + C_{ji}) \ , \forall j = 1, ..., N_G \ ,
\end{equation}
where $N_G$ is the number of sub-systems.

Finally, for this work we consider the third order dynamic system presented in \eqref{eq:NL_model_for_simulation} which implies $N_G = 3$ and therefore the overall TS system is represented as
\begin{equation} \label{eq:complete_polytopic_TS_sys}
    x^+ = \sum_{i=1}^{N_v} 
    \left[\begin{array}{c}
     	\mu_{N_{1i}}  \\
     	\mu_{N_{2i}}  \\
     	\mu_{N_{3i}}  
    \end{array} \right] \Bigg( 
      	\left[\begin{array}{c}
     	A_{1i} \\
     	A_{2i} \\
     	A_{3i}
    \end{array} \right] x +
      	\left[\begin{array}{c}
     	B_{1i} \\
     	B_{2i} \\
     	B_{3i}
    \end{array} \right] u +    	
      	\left[\begin{array}{c}
     	C_{1i} \\
     	C_{2i} \\
     	C_{3i}
    \end{array} \right] \Bigg) \ .
\end{equation} 
From now on, with the aim of an easier reading, the system representation in \eqref{eq:complete_polytopic_TS_sys} will be expressed as
\begin{equation}
	\label{eq:final_TS_system}
	x_{k+1} = \sum \limits_{i = 1}^{N_v} \mu_{N_{i}}(\zeta_k) (A_{i} x_k + B_{i} u_k + C_{i}) \ .
\end{equation}

\section{TS Control and Estimation} \label{section:4}
In this section, we present the formulations for the MPC and MHE techniques using the TS formulation.
\subsection{TS-MPC Design} \label{section:4.1}
Computing the predicting states behaviour in a certain horizon when using a system dependent on some scheduling variables (TS system) can be a challenging task sometimes leading to errors in the instantiation since the real future behaviour is unknown.

In this work, we propose the use of data coming from two different locations to approximate in a better way the predictive instantiation and avoid a lack of convergence in the optimal procedure.
On the one hand, data coming from a planner is used which represents the desired states behaviour for tracking the desired trajectory.
On the other hand, predicted states from the past optimal realisation are also used to improve the TS model instantiation.
\begin{figure}
	\centering
    \includegraphics[width=80mm]{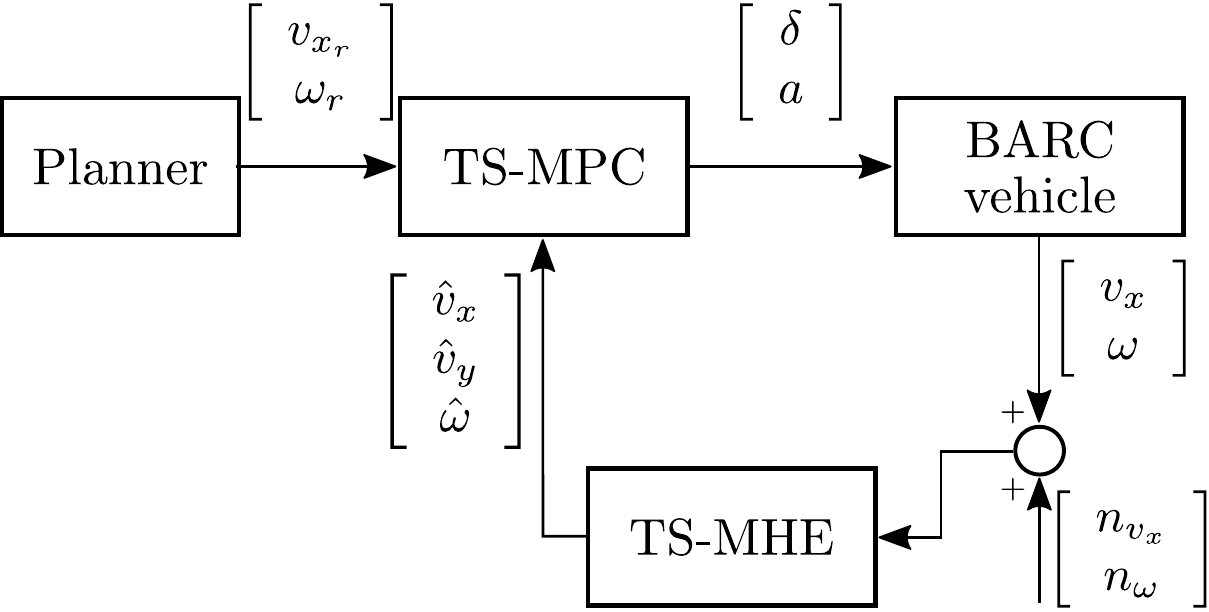}
    \caption{Schematical view of the simulation set up}
    \label{fig:schematical_view}
\end{figure}
The model used in this section is the one presented in \eqref{eq:final_TS_system} where the vector of scheduling variables is defined as $\zeta := \left[\begin{array}{ccccc} v_x & v_y & \omega & \delta & a \end{array}\right]$.
The use of this model allows to formulate the MPC problem as a quadratic optimization problem that is solved at each time $k$ to determine the control actions considering that the values of $x_k$ and $u_{k-1}$ are known
\begin{equation}
    \label{eq:MPC}
    \begin{aligned}
    & \underset{\Delta U_k}{\text{min}}
    && \mathrm{J_k}= \sum_{i=0}^{H_p-1} \Big( (r_{k+i} - x_{k+i})^T Q (r_{k+i} - x_{k+i}) \\
    &&& + \Delta u_{k+i} R \Delta u_{k+i} \Big) + x_{k+H_p}^T P x_{k+H_p} \\
    & \text{s.t.}
    && x_{k+i+1} = \sum \limits_{j = 1}^{N_v} \mu_{N_{j}}(\zeta_k) (A_{j} \hat{x}_{k+i} + B_{j} u_{k+i} + C_{j}) \\
    &&& u_{k+i} = u_{k+i-1} + \Delta u_{k+i} \\
    &&& \Delta U_k \in \Delta \Pi \\
    &&& U_k \in \Pi  \\
    &&& x_{k+H_p} \in \chi \\
    &&& y_{e} \in [\underline{y_e}, \overline{y_e} ] \\
    &&& x_{k+0} = \hat{x}_k   \ , \
    \end{aligned}
\end{equation}
where $\Pi = \{ u_k | A_u u_k = b_u , u_k \ge 0 \}$ and $\Delta \Pi = \{ \Delta u_k | A_{\Delta u} \Delta u_k = b_{\Delta u} , \Delta u_k \ge 0 \}$ constraint the system inputs and their variations, respectively.

State vector is $x = \left[\begin{array}{ccc} v_x & v_y & \omega \end{array}\right]^T$, $\hat{x}$ is the estimated state vector, $r = \left[\begin{array}{ccc} v_{x_{r}} & 0 & \omega_r \end{array}\right]^T$ is the reference vector provided by the trajectory planner, $u = \left[\begin{array}{cc}\delta & a \end{array}\right]^T$ is the control input vector and $H_p$ is the prediction horizon.
The tuning matrices $Q \in \mathbb{R}^{3x3}$ and $R \in \mathbb{R}^{2x2}$, are positive definite in order to obtain a convex cost function.
The closed loop stability is guaranteed by introducing $P \in \mathbb{R}^{3x3}$ and $\chi$ which represent the terminal set and the terminal constraint, respectively. Both are computed following the design presented in \cite{alcala2019ts}.
Note that the time discretization is embedded inside the identification procedure such that the learned TS system is already in discrete time.

\subsection{TS-MHE Design} \label{section:4.2}
For the vehicle presented in Section 2, vehicle lateral velocity ($v_y$) is an unmeasurable variable and a necessary state to perform the closed-loop control of the vehicle.
In this paper, we solve the estimation problem using the MHE approach.
The aim of the MHE is to compute the current dynamic states by means of running a constrained optimization, using a set of past data and employing a system model for computing the current state.
At this point, using the presented TS model in \eqref{eq:final_TS_system}, we can run a quadratic optimization similar than in TS-MPC algorithm for estimating the current dynamic states as follows
\begin{equation}
    \label{eq:MHE}
    \begin{aligned}
    & \underset{\hat{X}_k}{\text{min}}
    && \mathrm{J_k} = \sum_{i=-H_p}^{0} \big( w_{k+i}^T Q w_{k+i} + s_{k+i}^T R s_{k+i} \big),  \\
    & \text{s.t.}
    && \hat{x}_{k+i+1} = \sum \limits_{j = 1}^{N_v} \mu_{N_{j}}(\zeta_k) (A_{j} \hat{x}_{k+i} + B_{j} u_{k+i} + C_{j}) \\
    &&& + w_{k+i} \\
    &&& y_{k+i} = C \hat{x}_{k+i} + s_{k+i} \\ 
    &&& \hat{X}_k \in X_d \\
    &&& \forall i=-H_p,...,0 \ , \ \\
    \end{aligned}
\end{equation}
that is solved online for
\begin{equation}
        \hat{X}_k = \left[\begin{array}{c}
                \hat{x}_{k-H_p+1}     \\
                \hat{x}_{k-H_p+2} \\
                \vdots  \\
                \hat{x}_{k+1} \\
    \end{array}\right]    \in \mathbb{R}^{H_p \times s} \ ,
\end{equation}
where $X_d$ is the constraint region for the dynamic states and its defined as $X_d = \{ x_k | A_x x_k = b_x , x_k \ge 0 \}$.
$H_p$ stands for the past data horizon and $s$ the number of states.
Matrices $Q = Q^T \in \mathbb{R}^{3x3}$ and $R = R^T \in \mathbb{R}^{2x2}$, are positive definite to generate a convex cost function and $w$ and $s$ represent the error of estimation and the process noise, respectively.
State and input vectors are $\hat{x} = \left[\begin{array}{ccc} v_x & v_y & \omega \end{array}\right]^T$ and $u = \left[\begin{array}{cc}\delta & a \end{array}\right]^T$.
Note that, unlike MPC technique, MHE strategy performs an optimization taken into account a window of past vehicle data.

\section{Results} \label{section:5}

\begin{figure}[h]
	\centering
    \includegraphics[width=70mm]{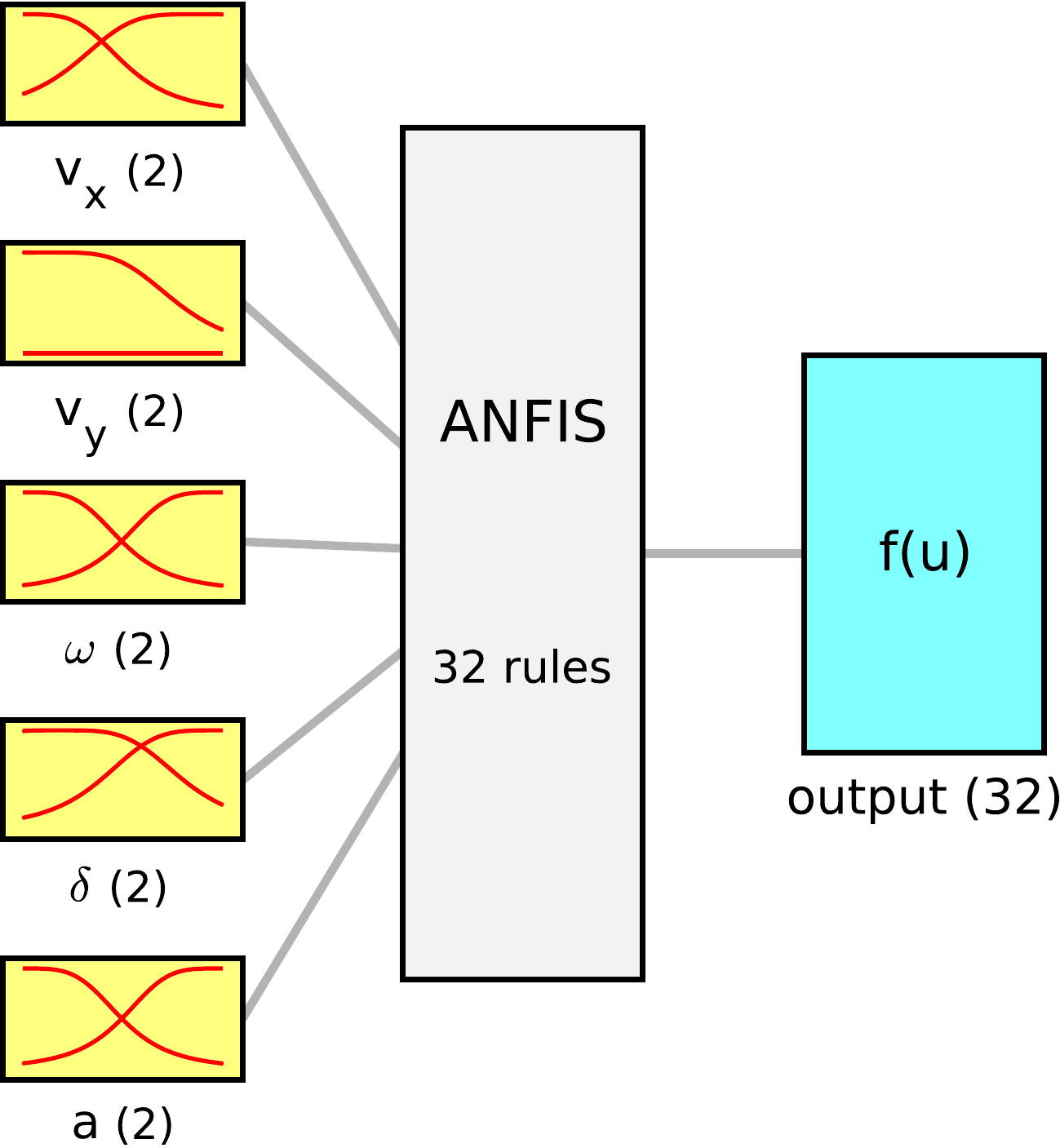}
    \caption{Input-Output scheme for the $v_x$ sub-system case}
   	\label{fig:vx_anfis_scheme}
\end{figure}

%% Primero hablar de la identificacion usando ANFIS
The data-driven model identification carried out by the proposed approach is used to learn a state-space TS formulation of the real vehicle dynamics.
In Figure \ref{fig:vx_anfis_scheme}, the membership functions learned for each input after the offline identification procedure are shown in the left side.
These represent the fuzzy rules that will be used online for obtaining the current state-space representation.

Note that, since the discretization time is embedded in the input data of the learning procedure, the selection of a different sampling time is not allowed.

%% Luego del set-up del MPC
The way of evaluating the MPC and MHE strategies using the data-driven approach presented is through a simulation scenario. In this, a racing situation is proposed where the autonomous scheme presented in Figure \ref{fig:schematical_view} is simulated. 

First, at every sampling period, i.e. 30 Hz, the racing planner provides the references for the control strategy such that the vehicle will have to behave in a racing driving mode, what directly implies a more challenging control problem.
Then, the TS-MHE optimal problem presented in \eqref{eq:MHE} is solved for estimating the current vehicle vector state using past vehicle measurements.
The next step is to instantiate the TS model matrices for the prediction stage using the approach presented in Section \ref{section:2}.
Note that, both the planning evolution information as the previous optimal prediction are used to achieve a good guess of the future values of the scheduling vector $\zeta$.
At this point, the quadratic optimal problem \eqref{eq:MPC} is solved using the estimated state variables and the references coming from the trajectory planner.
Once the optimal control actions ($\delta$ and $a$) are computed they are applied to the simulation vehicle presented in \eqref{eq:NL_model_for_simulation}.
As a consequence, the vehicle change its state and this is measured by the net of sensors.
Besides, with the aim of adding more realistic conditions to the problem, white Gaussian noise magnitudes are added to measured states with zero mean and covariances
\begin{equation}
	\begin{aligned}
	& Co_{v_x} = 1 \times 10^{-6} \ , \ Co_{\omega} = 4 \times 10^{-8} \ .
	\end{aligned}
\end{equation}

Both TS-MPC and TS-MHE algorithms are coded in MATLAB framework. Yalmip and GUROBI \citep{gurobi2015gurobi} are used for solving a quadratic optimization problem and running on a DELL inspiron 15 (Intel core i7-8550U CPU @ 1.80GHzx8).
In the controller, the tuning aims to minimize the longitudinal and angular velocity while computing smooth control actions.
The diagonal terms of the weighting matrices in the cost function and prediction horizon of \eqref{eq:MPC}, found by iterative tuning until the desired performance is achieved, are
\begin{equation}
    \begin{aligned}
    & Q = 0.65 \ [\begin{array}{ccc}
         0.4 & 10^{-6} & 0.6
    \end{array}], \\
    & R = 0.35 \ [\begin{array}{cc}
         0.7 & 0.3
    \end{array}],  \\
    & H_p = 6 \ .
    \end{aligned}
\end{equation}
The TS-MPC input constraints are defined as
\begin{subequations}
    \begin{equation}
        A_{u} = \left[\begin{array}{cc}
         	1 	& 0 \\
         	-1	& 0 \\
         	0 	& 1 \\
         	0 	& -1
    	\end{array} \right] , \
       	b_{u} = \left[\begin{array}{c}
         	0.249 \\
         	0.249 \\
         	4 \\
         	1
    	\end{array} \right] , \
    \end{equation}
    \begin{equation}
        A_{\Delta u} = \left[\begin{array}{cc}
         	1 	& 0 \\
         	-1	& 0 \\
         	0 	& 1 \\
         	0 	& -1
    	\end{array} \right] , \
       	b_{\Delta u} = \left[\begin{array}{c}
         0.05 \\
         0.05 \\
         0.5 \\
         0.5
    	\end{array} \right] \ .
    \end{equation}
\end{subequations}

In the estimator, the tuning aims to minimize the process noise while guessing the right value of $v_y$ by using the TS model.
The diagonal terms of the weighting matrices in the cost function and past horizon of \eqref{eq:MHE}, found by iterative tuning until the desired performance is achieved, are
\begin{equation}
    \begin{aligned}
    & Q = 0.5 \ [\begin{array}{ccc}
         0.25 & 0.5 & 0.25
    \end{array}], \\
    & R = 0.5 \ [\begin{array}{cc}
         0.5 & 0.5
    \end{array}],  \\
    & H_p = 10 \ .
    \end{aligned}
\end{equation}
The TS-MHE state region is defined in the polytope
\begin{subequations}
    \begin{equation}
        A_{x} = \left[\begin{array}{ccc}
         	1 	& 0  & 0 \\
         	-1	& 0  & 0 \\
         	0 	& 1  & 0 \\
         	0 	& -1 & 0 \\
         	0 	& 0  & 1 \\
         	0 	& 0  & -1
    	\end{array} \right] , \
       	b_{x} = \left[\begin{array}{c}
         	2.7 \\
         	-0.1 \\
         	0.12 \\
         	0.12 \\
         	1.96 \\
         	1.96 
    	\end{array} \right] .
    \end{equation}
\end{subequations}

\begin{figure}[htpb!]
	\centering
    \includegraphics[width=95mm]{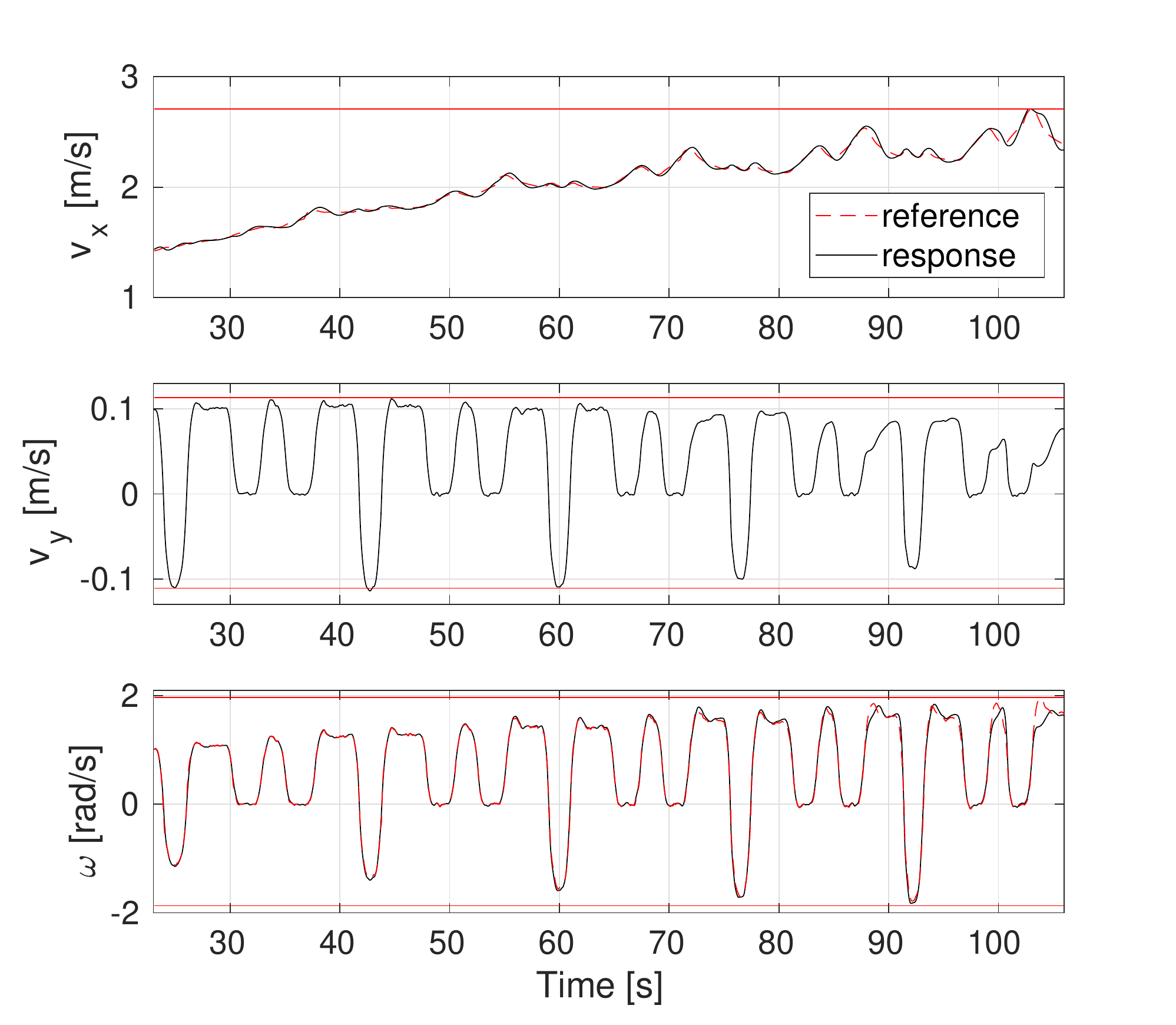}
    \caption{Vehicle states throughout the simulation. Horizontal red lines represent the upper and lower limits}
   	\label{fig:simu_velocity}
\end{figure}
Figure \ref{fig:simu_velocity} shows both the reference and the response for each one of the velocity variables.
Note that, the vehicle lateral velocity ($v_y$) cannot be measured and hence, the signal presented is the estimated one.
It can be seen that the controller is able to perfectly track the proposed references although having little troubles when driving in racing mode, i.e. after 85 seconds.
Horizontal red lines represent the polytope boundaries for each one of the scheduling variables that in this approach coincide with the state and input vehicle variables. Note that, this limits are imposed in the learning stage by the maximum and minimum values of the input signals, i.e. scheduling variables.

\begin{figure}[htpb!]
    \centering
    \includegraphics[width=95mm]{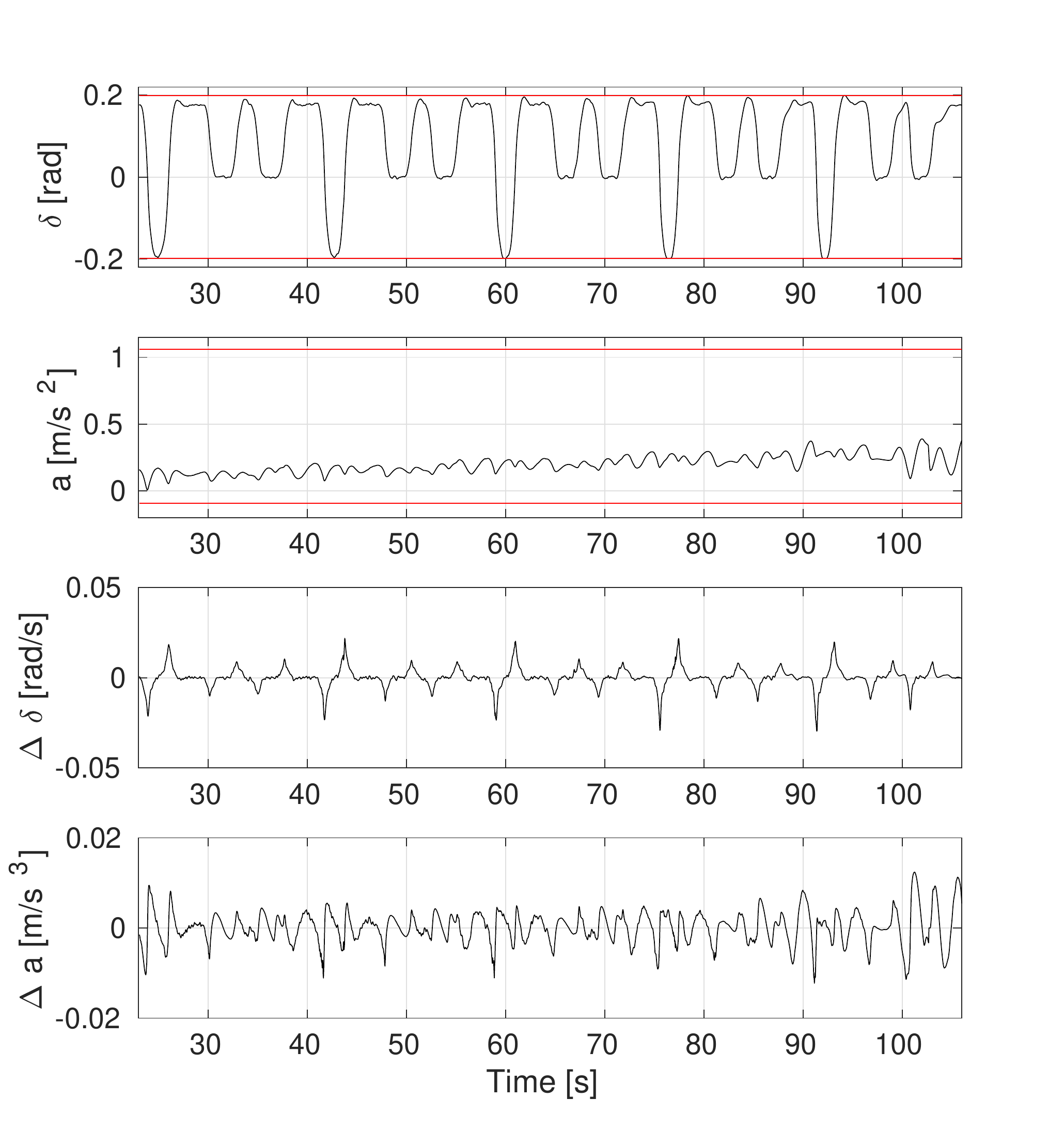}
    \caption{Control actions and their time derivative variables throughout the simulation. Horizontal red lines represent the upper and lower limits}
   	\label{fig:control_actions}
\end{figure}

In Figure \ref{fig:control_actions}, the optimal control actions are shown as well as their discrete time variations which are the ones minimized in the cost function of \eqref{eq:MPC}. Note that the steering angle reaches the upper and lower limits at some points while the rear wheel acceleration moves in a wider range.

\begin{figure}[htpb!]
    \centering
    \includegraphics[width=95mm]{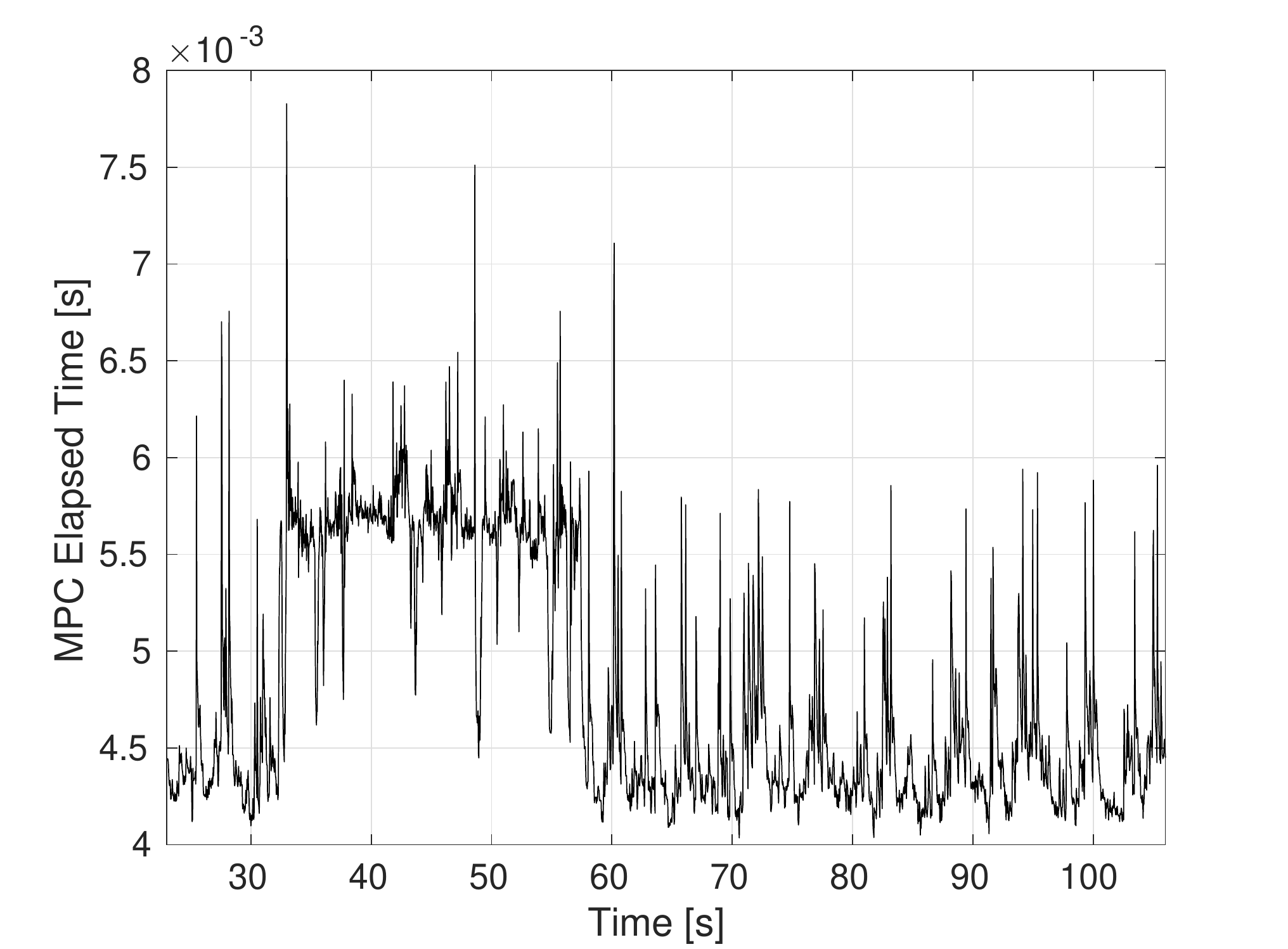}
    \caption{Computational time required by the TS-MPC throughout the simulation}
   	\label{fig:elapsed_time_MPC}
\end{figure}

Finally, after observing a good tracking performance in last figures, we present the elapsed time per iteration of the TS-MPC in Figure \ref{fig:elapsed_time_MPC}. 
It can be seen that, using a prediction horizon of 6 steps, the quadratic solver is able to obtain an average of 4.8 ms. This is one of the most remarkable results of this approach.

\section{Conclusions} \label{section:6}
In this paper, a learning-based approach for identifying the dynamics of the vehicle and formulating them as a TS representation has been presented.
Then, a TS-MPC strategy has been proposed as the approach to solve autonomous driving control problems under realistic conditions in real-time.
In addition, using racing-based references provided by an external planner the controller makes the vehicle to perform in racing mode.
The control strategy has been tested in simulation showing high performance potential in both reference tracking and computational time. 
However, this approach share the limitation of learning-based procedures where you can only do what you learn.

\begin{ack}
This work has been funded by the Spanish Ministry of Economy and Competitiveness (MINECO) and FEDER through the projects SCAV (ref. DPI2017-88403-R) and HARCRICS (ref. DPI2014-58104-R). 
The author is supported by a FI AGAUR grant (ref 2017 FI B00433).
\end{ack}

\bibliography{ifacconf}             % bib file to produce the bibliography

\end{document}